\documentclass[aip,amsmath,amssymb,reprint]{revtex4-1}

\usepackage{graphicx}
\usepackage{dcolumn}
\usepackage{bm}

\usepackage[utf8]{inputenc}
\usepackage[T1]{fontenc}
\usepackage{mathptmx}
\usepackage{etoolbox}

\usepackage{graphicx}
\usepackage{dcolumn}
\usepackage{bm}
\usepackage{mathrsfs}
\usepackage{amsmath,gensymb}
\usepackage{amsfonts}
\usepackage{amssymb}
\usepackage{amsthm}
\usepackage{graphicx}
\usepackage{natbib}
\usepackage{xcolor}
\usepackage{hyperref}
\hypersetup{
    colorlinks=true,
    linkcolor=blue,
    filecolor=magenta, 
    citecolor=blue,
    urlcolor=blue,
}
\usepackage{bm}
\usepackage[caption=false]{subfig}
\usepackage{verbatim}
\usepackage[utf8]{inputenc}
\usepackage{soul}
\usepackage{braket}
\usepackage{comment}
\usepackage{ulem}
\makeatletter
\def\@email#1#2{%
 \endgroup
 \patchcmd{\titleblock@produce}
  {\frontmatter@RRAPformat}
  {\frontmatter@RRAPformat{\produce@RRAP{*#1\href{mailto:#2}{#2}}}\frontmatter@RRAPformat}
  {}{}
}%
\makeatother

\newcommand{\showcomments}{\long\def\comm##1\commend{##1}}

\showcomments

\newcommand{\simcom}[1]{\comm \textcolor{green}{[{#1}]} \commend}

\begin{document}

\preprint{AIP/123-QED}
	
\title{Thermodynamic properties of a superconductor interfaced with an altermagnet}

\author{Simran Chourasia}
\email{simran.chourasia@uam.es,wolfgang.belzig@uni-konstanz.de}
\affiliation{Condensed Matter Physics Center (IFIMAC) and Departamento de F\'{i}sica Te\'{o}rica de la Materia Condensada, Universidad Aut\'{o}noma de Madrid, E-28049 Madrid, Spain}

\author{Aleksandr Svetogorov}
\affiliation{Fachbereich Physik, Universit\"{a}t Konstanz, D-78457 Konstanz, Germany}

\author{Akashdeep Kamra}
\affiliation{Condensed Matter Physics Center (IFIMAC) and Departamento de F\'{i}sica Te\'{o}rica de la Materia Condensada, Universidad Aut\'{o}noma de Madrid, E-28049 Madrid, Spain}

\author{Wolfgang Belzig}
\affiliation{Fachbereich Physik, Universit\"{a}t Konstanz, D-78457 Konstanz, Germany}

\date{\today}

\begin{abstract}
    Recently introduced magnetic materials called altermagnets (AM) feature zero net magnetization but a momentum-dependent magnetic exchange field, which can have intriguing implications when combined with superconductivity. In our work, we use the quasiclassical framework to study the effects of such a material on a conventional superconductor (S) in an AM/S bilayer. We discuss the superconducting phase diagram and heat capacity of AM/S while making a comparison with a ferromagnet-superconductor bilayer. Furthermore, we examine the density of states and analyze the system's response to an external magnetic field. We illustrate the anisotropy of spin-susceptibility and magnetization of AM/S by considering an external field in the in-plane and out-of-plane direction, thereby facilitating the scope of experimental detection and characterization of an AM in an AM/S hybrid system.
\end{abstract}

\maketitle

The interplay of superconductivity and magnetism has been receiving significant attention in condensed matter physics for decades~\cite{Bergeret2005, Bergeret2018, Buzdin2005, Linder2015, Eschrig2008}. Nevertheless, the discoveries of novel hybrid materials and new experimental techniques constantly provide new interesting phenomena to study~\cite{Baek2014, Khaire2010, Jeon2018, Jeon2020, Diesch2018, Chiodi2013, Wu2012}. One of the latest additions to magnetic materials is a collinear symmetry-compensated antiferromagnet~\cite{Hayami2019, Yuan2020, Mazin2021}, for which the term "altermagnet" was introduced~\cite{Jungwirth2022_1, Jungwirth2022_2, Mazin2022Editorial, Mazin2024Physics}. Recently, experimental evidence for that new type of magnetism has been reported~\cite{Kim2024}. Intrinsic anomalous Hall effect has also been predicted~\cite{Attias2024}. The idea to combine this new type of material with superconductors emerged~\cite{Mazin2022notes} right after theoretical predictions of altermagnets. Andreev reflection~\cite{Sun2023, Wei2023, Papaj2023, Nagae2024}, Josephson effect~\cite{Ouassou2023, Beenakker2023, Zhang2023}, superconducting diode effect~\cite{Banerjee2024} and topological superconductivity~\cite{Li2023, Ghorashi2023} were studied theoretically in altermagnet-superconductor (AM/S) hybrid structures among others \cite{zyuzin2024magnetoelectric,giil2024quasiclassical}. It is known that both a ferromagnet~\cite{Tedrow1986, Buzdin2005, Jiang1996}, which has a spin-split band, as well as an antiferromagnet~\cite{Hubener2002, Bobkov2022, Kamra2023} with spin-degenerate band modify the properties of a superconductor when placed in its proximity. In this work, we investigate the effects an altermagnet, which has a spin-split band as well as zero net magnetization, has on a superconductor in a thin film AM/S bilayer. Our objective is to study the magnetic proximity effect that results in properties that can be experimentally measured and distinguish an AM/S bilayer from bilayers incorporating other magnetic materials.

By formulating the model in the quasiclassical framework, we find that the spin-resolved density of states of the superconductor in an AM/S bilayer has distinct features, and can lead to gapless superconductivity for a strong AM. We also study the superconducting phase diagram of the AM/S bilayer and find that the superconductor to normal metal (N) phase transition with increasing temperature is always second-order. Further, we find that spin susceptibility and magnetization of AM/S in the presence of an external magnetic field show unique trends which can act as an experimental signature of d-wave altermagnetism.

\begin{figure*}[tb]
    \includegraphics[width=\linewidth]{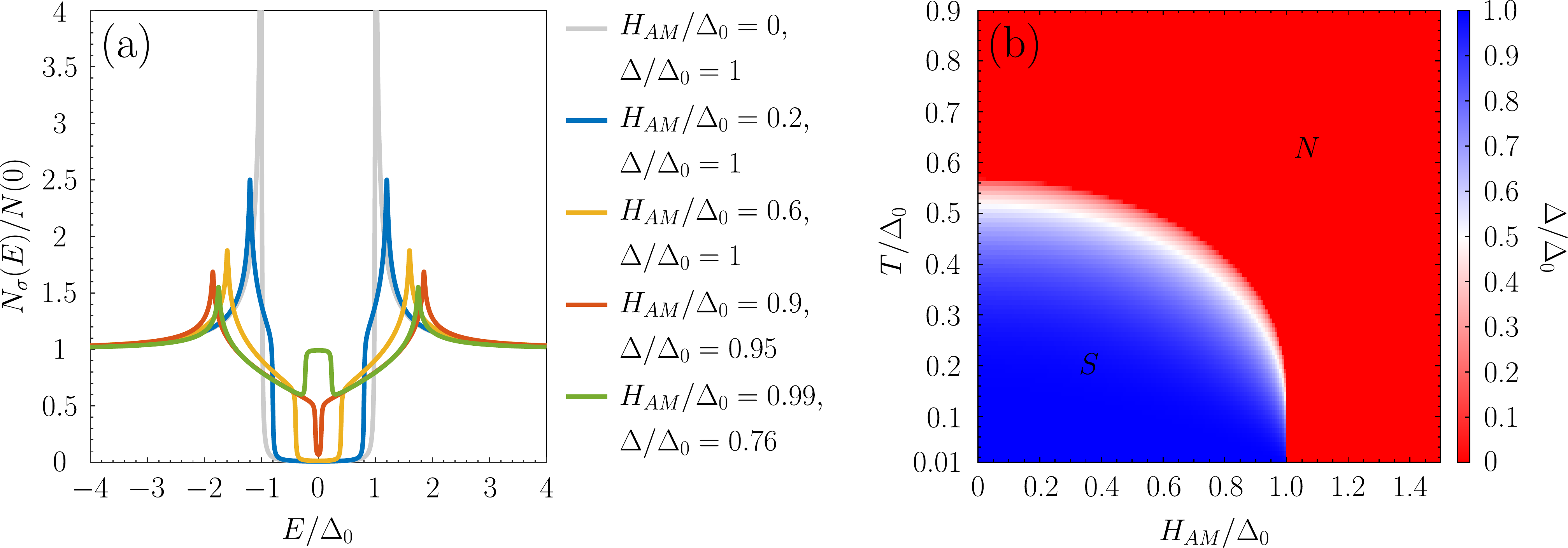}
    \caption{(a) Superconductor's density of states (DOS) in an altermagnet-superconductor bilayer (AM/S) for different exchange strengths $H_{AM}$ of the altermagnet is plotted for spin $\sigma$. Since the DOS of AM/S is found same for both the spins, the curve for only spin-$\uparrow$ quasiparticles is shown. Here, the DOS is plotted at temperature $T=0.1\,\Delta_0$, $N(0)$ is the DOS of a normal metal for each spin, the gray curve is the DOS of an isolated superconductor, and $\Delta_0$ is the superconducting order parameter of an isolated superconductor at zero-temperature. (b) Phase diagram of AM/S obtained by plotting its superconducting order parameter $\Delta$ for different values of $T$ and $H_{AM}$. The blue and red regions correspond to superconducting and normal states of the AM/S, respectively.}
    \label{plot_DOS_and_PhaseDiagram}
\end{figure*}

\begin{figure*}[tb]
    \includegraphics[width=\linewidth]{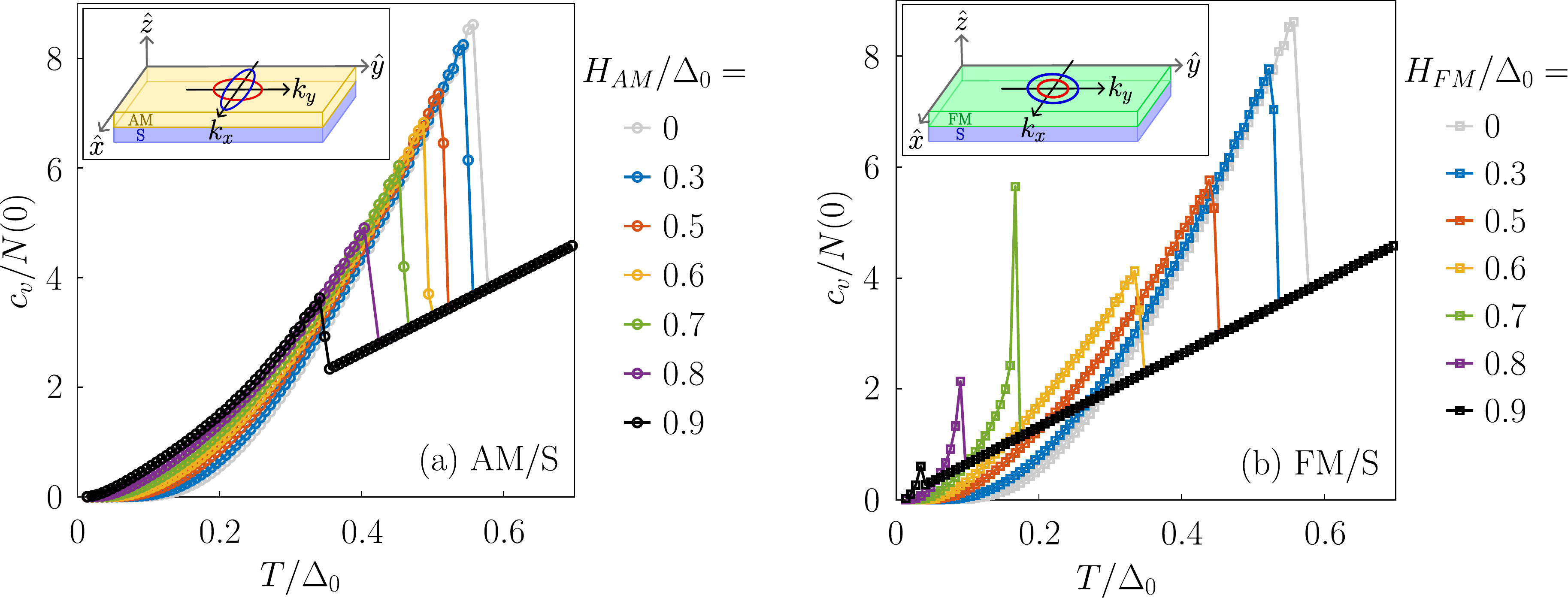}
    \caption{Superconductor's temperature dependence of heat capacity per unit volume $c_v$ in (a) altermagnet-superconductor (AM/S), and (b) ferromagnet-superconductor (FM/S) bilayers for different values of exchange field strengths. The insets show the schematic diagrams of the corresponding (a) AM/S and (b) FM/S bilayers with their normal-state Fermi surfaces of spin-up (blue) and -down (red) electrons drawn in the $k_x$-$k_y$ plane.}
    \label{SpecificHeatCapacity_final}
\end{figure*}

Consider a thin film of a conventional s-wave superconductor (S) interfaced with an insulating altermagnet (AM) such that the thickness of the superconductor is much smaller than its coherence length. The AM affects the superconductor by imparting a momentum-dependent magnetic exchange field on the electrons of the superconductor [Fig.~\ref{SpecificHeatCapacity_final}(a) inset]~\cite{Papaj2023, Beenakker2023}. Using the quasiclassical framework, the conduction electrons of the superconductor can be described by the following Eilenberger equation~\cite{Eilenberger1968}
\begin{align} \label{Eilenberger1}
    \left[ \omega_n (\bm{\hat{\sigma}_0} \otimes \bm{\hat{\tau}_z}) + \bm{\check{\Delta}} - i \, \vec{H}_{\text{eff}}(\hat{k}) \cdot \bm{\hat{\vec{\sigma}}} \otimes \bm{\hat{\tau}_z}, \, \bm{\check{g}_{\omega_n}}(\hat{k})\right]=\bm{\check{0}},
\end{align}
where $\omega_n=(2n+1)\pi T$ is a Matsubara frequency for an integer $n$, $T$ is the temperature, $\hat{k}=(\sin \theta \cos \phi \, \hat{x} + \sin \theta \sin \phi \, \hat{y} + \cos \theta \, \hat{z})$ is the unit vector along the crystal momentum $\vec{k}$ of the electron on the Fermi surface being described, and $\theta$ and $\phi$ are the angles describing $\hat{k}$ in the spherical coordinates. Matrices $\bm{\hat{\sigma}_0}$ ($\bm{\hat{\tau}_0}$) and $\bm{\hat{\sigma}_\alpha}$ ($\bm{\hat{\tau}_\alpha}$) with $\alpha \in \{x,y,z\}$ are identity and Pauli matrices in the spin (Nambu) space. The superconducting pairing is described by $\bm{\check{\Delta}} = \Delta (\bm{\hat{\sigma}_0} \otimes \bm{\hat{\tau}_x})$, where $\Delta$ (assumed real) is the s-wave superconducting order parameter. The electrons of the superconductor experience an effective d-wave exchange field, $\vec{H}_{\text{eff}}(\hat{k}) \equiv \vec{H}_{\text{eff}}(\phi)=H_{AM} \cos (2\phi) \hat{z}$, due to its proximity to the AM~\cite{Beenakker2023, Papaj2023}. This means the energies of spin-up and -down electrons with wavevector $\vec{k}$ get shifted by $\mp H_{AM} \cos (2\phi)$. Here, $H_{AM}$ is the amplitude of the altermagnetic exchange field and a higher $H_{AM}$ means a stronger altermagnet. The quasi-classical Green's function $\bm{\check{g}_{\omega_n}}$, written in spin $\otimes$ Nambu space, contains all the relevant information of the system~\cite{Kopnin2001}. By satisfying Tr $\bm{\check{g}_{\omega_n}}=0$ and $\bm{\check{g}_{\omega_n}}\bm{\check{g}_{\omega_n}}=\bm{\check{1}}$, we get
\begin{align}
    & \bm{\check{g}_{\omega_n}}(\hat{k})= 
    \begin{bmatrix}
        g_{\uparrow \uparrow,\omega_n} & if_{\uparrow \downarrow,\omega_n} & g_{\uparrow \downarrow,\omega_n} & i f_{\uparrow \uparrow,\omega_n}\\
        i f^\dagger_{\downarrow \uparrow,\omega_n} & \bar{g}_{\downarrow \downarrow,\omega_n} & i f_{\downarrow \downarrow,\omega_n}^{\dagger}& \bar{g}_{\downarrow \uparrow,\omega_n} \\
        g_{\downarrow \uparrow,\omega_n} & i f_{\downarrow \downarrow,\omega_n} & g_{\downarrow \downarrow,\omega_n} & i f_{\downarrow \uparrow,\omega_n} \\
        i f^\dagger_{\uparrow \uparrow,\omega_n} & \bar{g}_{\uparrow \downarrow,\omega_n} & i f_{\uparrow \downarrow,\omega_n}^\dagger & \bar{g}_{\uparrow \uparrow,\omega_n}
    \end{bmatrix} \label{gGeneralExpression}\\
    & = \frac{1}{2} \left(\bm{\hat{\sigma}_0}+\bm{\hat{\sigma}_z}\right) \otimes \bm{\hat{g}_{+,\omega_n}}(\phi) + \frac{1}{2} \left(\bm{\hat{\sigma}_0}-\bm{\hat{\sigma}_z}\right) \otimes \bm{\hat{g}_{-,\omega_n}}(\phi), \nonumber
\end{align}
with
\begin{align}
    \bm{\hat{g}_{\pm,\omega_n}}(\phi) = \frac{\left[(\omega_n \mp i \, H_{AM} \cos (2\phi))\bm{\hat{\tau}_z}+\Delta \bm{\hat{\tau}_x} \right]}{\sqrt{(\omega_n \mp i \, H_{AM} \cos (2\phi))^2+ \Delta^2}}. 
\end{align}
The order parameter $\Delta$ is calculated self-consistently as~\cite{Kopnin2001}
\begin{align} \label{SelfConsistentDeltaAMS_1}
    \Delta = i \pi \lambda T \sum_{n=0}^{N_0} \int_0^{2\pi} \frac{d\phi}{2\pi} \left[ f_{\uparrow \downarrow, \omega_n}+f_{\downarrow \uparrow, \omega_n} \right],
\end{align}
where $\lambda$ is the interaction constant, $N_0=\lceil (\Omega_{BCS}/\pi T -1)/2 \rceil$ depends on the BCS cutoff frequency $\Omega_{BCS}=\frac{1}{2} \Delta_0 \, e^{1/\lambda}$, and $\Delta_0$ is the superconducting order parameter of an isolated superconductor at zero temperature. 

Using this model, we describe how the AM modifies properties of the superconductor in an AM/S bilayer. The exchange field of the AM is assumed to remain constant for temperatures and magnetic fields varying within an order of $\Delta_0$. Therefore, in this letter, AM/S represents the superconducting layer of the AM/S bilayer. 

The density of states (DOS), $N_\sigma(E)$, of quasiparticles with energy $E$ and spin $\sigma$ in the AM/S can be calculated, by replacing $i \omega_n$ with $E+i\eta$ in Eq.~\ref{gGeneralExpression} of $\bm{\check{g}_{\omega_n}}(\hat{k})$, as
\begin{align}
    \frac{N_\sigma(E)}{N(0)} = \int_{0}^{2\pi} \frac{d\phi}{2\pi} \text{Re} \left[ g_{\sigma \sigma,\omega_n} \right]_{i \omega_n\rightarrow E+i\eta},
\end{align}
where $\eta$ is a small positive number and $N(0)$ is the DOS of each spin in a normal metal. Fig.~\ref{plot_DOS_and_PhaseDiagram}(a) shows the DOS of AM/S for different strengths of altermagnetic exchange field. In AM/S, the change in the sign of $\vec{H}_{\text{eff}}(\hat{k})$ with changing $\phi$ in the $\vec{k}$-space ensures that the DOS of spin-$\uparrow$ and $\downarrow$ quasiparticles are same at each energy. We find that $N_{\sigma}(E)/N(0)$ has peaks at $E=\pm|\Delta + H_{AM}|$ with logarithmic divergence and shoulder-like structures at $E=\pm|\Delta-H_{AM}|$ for $H_{AM}<\Delta$. For $H_{AM}>\Delta$, the two shoulder-like structures of the DOS overlap with each other and lead to the closing of the gap in the spectrum, while retaining superconductivity, as in the case for the green curve in Fig.~\ref{plot_DOS_and_PhaseDiagram}(a). The DOS evaluated for AM/S resembles the normalized total DOS, $[N_{\uparrow}(E)+N_{\downarrow}(E)]/2N(0)$, of the superconducting layer of an FM/S bilayer weighted averaged over the magnitude of exchange field varying from $0$ to $H_{AM}$.
	
In order to compare the superconducting phase diagram of AM/S with that of FM/S, we calculate $\Delta$ self-consistently and plot it, in Fig.~\ref{plot_DOS_and_PhaseDiagram}(b), for different values of altermagnetic exchange strength $H_{AM}$ and temperature $T$. We observe that the S-N transition with increasing temperature is a second-order transition at all values of $H_{AM}$. This is unlike the phase diagram of an FM/S bilayer, where the S-N transition becomes first-order for larger exchange field strengths (see Appendix). A rigorous approach would be to obtain the phase diagram by comparing different free energies involved in the system~\cite{Sarma1963, Maki1964}.

The difference in the phase diagrams of AM/S and FM/S gets reflected in the heat capacity per unit volume $c_v$, which is given by the expression~\cite{Tinkham1996}
\begin{align}
    \frac{c_v}{N(0)} = T \frac{d}{dT} \left[ \frac{S_{v}}{N(0)} \right],
\end{align}
where $S_{v}$ is the entropy per unit volume calculated as
\begin{align}
    S_{v}(T) =&-\int dE (N_\uparrow(E)+N_\downarrow(E)) \left[ f(E) \ln f(E)\right. \nonumber \\ 
    & \left.+ (1-f(E)) \ln (1-f(E)) \right],
\end{align}
and $f(E)=1/(e^{E/T}+1)$ is the Fermi-Dirac distribution function with the Boltzmann constant set to 1. Fig.~\ref{SpecificHeatCapacity_final} shows the temperature dependence of $c_v/N(0)$ for AM/S [Fig.~\ref{SpecificHeatCapacity_final}(a)] and FM/S [Fig.~\ref{SpecificHeatCapacity_final}(b)]. The insets of Fig.~\ref{SpecificHeatCapacity_final}(a) and Fig.~\ref{SpecificHeatCapacity_final}(b) show the schematic diagrams of AM/S and FM/S bilayers. The calculations for FM/S are done by substituting $\vec{H}_{\text{eff}}(\hat{k})=H_{FM}\hat{z}$ in Eq.~\ref{Eilenberger1} and solving it. From Fig.~\ref{SpecificHeatCapacity_final}(b), we find that the change of S-N transition from second-order to first-order for larger $H_{FM}$ gets manifested as sharper peaks in $c_v/N(0)$ curves. On the other hand, the $c_v/N(0)$ versus $T$ curves of AM/S change gradually with increasing value of $H_{AM}$. Because the S-N transition in AM/S is second-order for all $H_{AM}$, we do not observe the appearance of sharper peaks in the $c_v$ curves with increasing exchange field in AM/S as we do in FM/S.


Since AM/S under study has a unique d-wave spin-splitting field, its magnetic response to an external magnetic field can help in probing it experimentally. To incorporate an external magnetic field $\vec{h}=h_x \, \hat{x} + h_y \, \hat{y} + h_z \, \hat{z}$ applied on AM/S, we substitute $\vec{H}_{\text{eff}}(\hat{k})=\vec{H}_{\text{eff}}(\phi)=\vec{h}+H_{AM} \cos (2\phi) \hat{z}$ in Eq.~\ref{Eilenberger1}. We then solve it for $\bm{\check{g}_{\omega_n}}(\hat{k})$ by satisfying Tr $\bm{\check{g}_{\omega_n}}=0$ and $\bm{\check{g}_{\omega_n}}\bm{\check{g}_{\omega_n}}=\bm{\check{1}}$, and using projector operators $\bm{\hat{P}_{\pm}}(\phi)$ in the spin-space for each $\vec{k}$ and find
\begin{align}
    & \bm{\check{g}_{\omega_n}}(\hat{k}) = \bm{\hat{P}_{+}}(\phi) \otimes \bm{\hat{g}_{+,\omega_n}}(\phi)+\bm{\hat{P}_{-,\omega_n}}(\phi) \otimes \bm{\hat{g}_{-,\omega_n}}(\phi),
\end{align}
where
\begin{align}
    \bm{\hat{g}_{\pm,\omega_n}}(\phi) = \frac{\left[(\omega_n \mp i |\vec{H}_{\text{eff}}(\phi)|)\bm{\hat{\tau}_z}+\Delta \bm{\hat{\tau}_x} \right]}{\sqrt{(\omega_n \mp i |\vec{H}_{\text{eff}}(\phi)|)^2+ \Delta^2}}
\end{align}
and $\bm{\hat{P}_{\pm}}(\phi) = \left[\bm{\hat{1}} \pm \vec{n}(\phi) \cdot \bm{\hat{\Vec{\sigma}}}\right]/2$ with $\Vec{n}(\phi)$ being the unit vector in direction of $\vec{H}_{\text{eff}}(\phi)$.
From $\bm{\check{g}_{\omega_n}}$, the magnetization carried by the conduction electrons is calculated as~\cite{Bernat_2023}
\begin{align}
    M_\alpha = M^N_\alpha + \frac{i}{4} \pi N(0) \mu_B g_\alpha T \sum_{\omega_n} \text{Tr} \, \langle (\bm{\hat{\sigma}_\alpha} \otimes \bm{\hat{\tau}_z}) \bm{\check{g}_{\omega_n}}(\hat{k}) \rangle_{\hat{k}},
\end{align}
where $\alpha \in \{x,y,z\}$, $M_\alpha$ is the magnetization along $\hat{\alpha}$-axis, $\mu_B$ is the Bohr magneton, $g_\alpha$ is the anisotropic Lande factor ($g_\alpha=-2$ for all $\alpha$ here), $\langle \rangle_{\hat{k}}$ means averaging over all directions in the $\vec{k}$-space, $M_\alpha^N=\chi_\alpha^N \langle \vec{H}_{\text{eff}}(\phi)\rangle_{\hat{k}}\cdot \hat{\alpha} = \chi_\alpha^N h_\alpha $ is the magnetization of the normal metal in an AM/N bilayer, and $\chi_\alpha^N = g_\alpha^2 \chi_P/4=\chi_P$ is the linear susceptibility of a normal metal, which is the same as Pauli susceptibility $\chi_P$ in our case.

As magnetization is an averaged quantity over all directions in the $\vec{k}$-space, it is found to be isotropic in the $x$-$y$ plane (in-plane) for AM/S [Fig.~\ref{SpecificHeatCapacity_final}(a) inset]. However, it is found to be different in the $\hat{z}$-direction (out-of-plane). We, therefore, study two cases: (a) when the external field is applied in-plane by choosing $\vec{h}=h \, \hat{x}$, and (b) when the external field is applied out-of-plane as $\vec{h}=h \, \hat{z}$. From now on, $\parallel$ represents that the applied field is in-plane and $\perp$ represents that the applied field is out-of-plane. 

\begin{figure}[tb]
    \includegraphics[width=\linewidth]{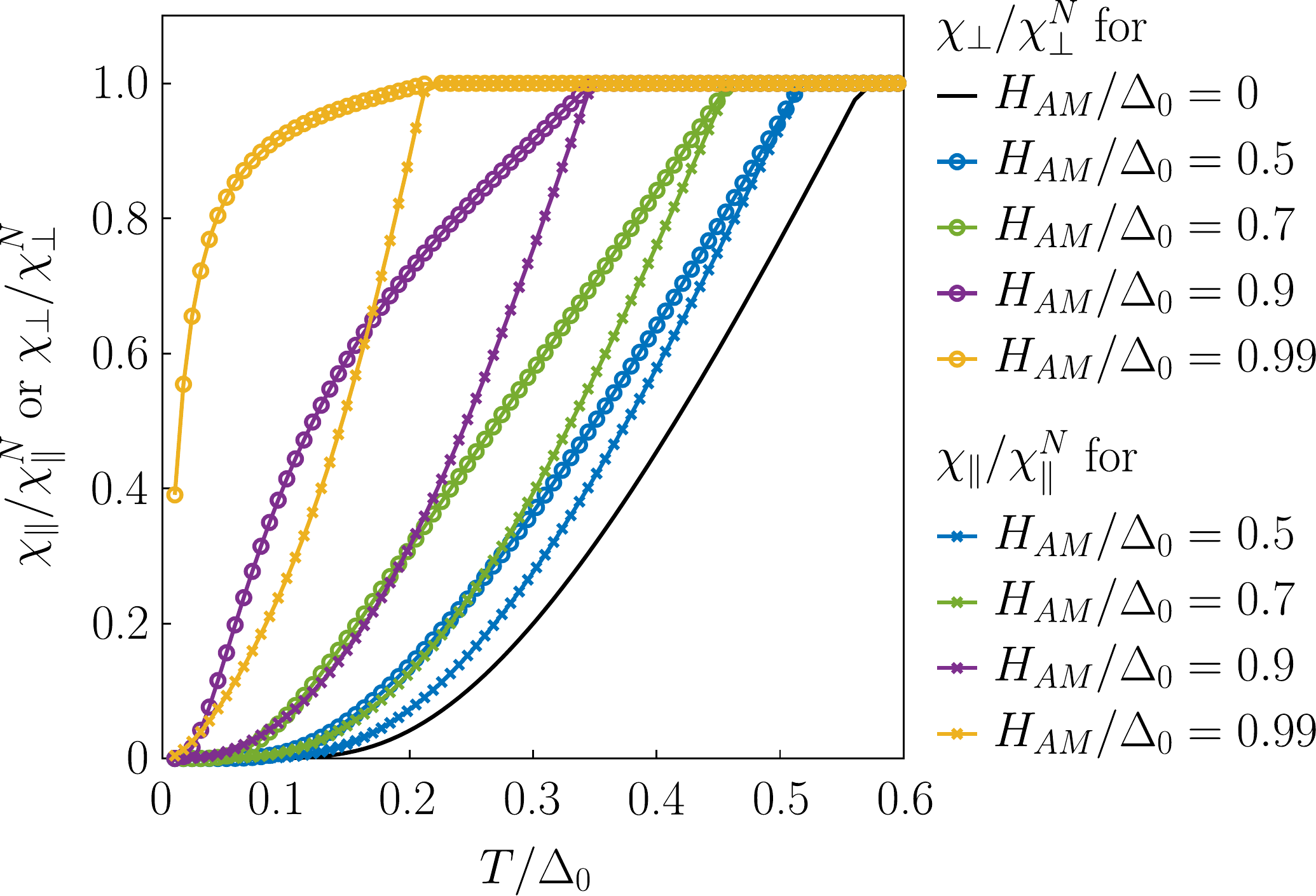}
    \caption{Superconductor's linear susceptibility in an AM/S bilayer is plotted as a function of temperature $T$ for different values of altermagnetic exchange strength $H_{AM}$. The curves with `x' and `o' markers correspond to the in-plane ($\chi_\parallel$) and out-of-plane ($\chi_\perp$) susceptibilities, respectively. Curves of the same colors correspond to the same AM/S parameters. Here, $\chi_\parallel^N$ and $\chi_\perp^N$ are the normal state values of the in-plane and out-of-plane susceptibilities, and an applied field of magnitude $h=0.01\Delta_0$ is used for the calculation.}
    \label{LinearSusceptibility_vs_T_AMS}
\end{figure}

We first study the linear susceptibility as it can be useful in characterizing the pairing symmetry of the AM/S. The in-plane (out-of-plane) linear susceptibility $\chi_{\parallel(\perp)}$ of the AM/S can be calculated from its magnetization in response to a very small magnetic field applied along $\hat{x}$~$(\hat{z})$-direction as
\begin{align} \label{SusceptibilityExpression}
    \frac{\chi_{\parallel(\perp)}}{\chi_{\parallel(\perp)}^N} = \lim_{h \to 0} \frac{M^{\parallel(\perp)}_{x(z)}}{M^{N,\parallel(\perp)}_{x(z)}},
\end{align}
where $\chi_{\parallel(\perp)}^N$ is the in-plane (out-of-plane) linear susceptibility and $M^{N,\parallel(\perp)}_{x(z)}$ is the magnetization along $\hat{x}$ ($\hat{z}$) for an applied in-plane (out-of-plane) magnetic field in an AM/N. The temperature dependence of linear susceptibilities of AM/S is shown in Fig.~\ref{LinearSusceptibility_vs_T_AMS}, where it is clear that $\chi_{\perp}/\chi_{\perp}^N \geq \chi_{\parallel}/\chi_{\parallel}^N$ at all temperatures. The in-plane susceptibility curves ($\chi_{\parallel}/\chi_{\parallel}^N$) are convex and similar in shape to that of an isolated superconductor (black curve in Fig.~\ref{LinearSusceptibility_vs_T_AMS}). Whereas the curves for the out-of-plane case ($\chi_{\perp}/\chi_{\perp}^N$), get modified from convex to roughly linear to highly concave curves with increasing $H_{AM}$. 

Since the intrinsic $\vec{k}$-dependent exchange field of the AM/S is in the out-of-plane direction, it does not have a direct effect on the in-plane susceptibility. However, AM's exchange field modifies the DOS of the superconductor in its proximity, leading to an indirect quantitative effect on $\chi_{\parallel}/\chi_{\parallel}^N$. This means that the $\chi_{\parallel}/\chi_{\parallel}^N$ curves for a superconductor get modified quantitatively and not qualitatively on interfacing it with an AM. On the other hand, an external field in the out-of-plane direction modifies the d-wave intrinsic exchange field of the AM/S such that the electrons experience stronger spin-splitting in the $k_x$ direction and a reduced magnitude in the $k_y$ direction. Applying the external field along $\hat{z}$ is therefore different in nature and allows one to probe the d-wave exchange field of the AM/S. 

\begin{figure}[tb]
    \includegraphics[width=\linewidth]{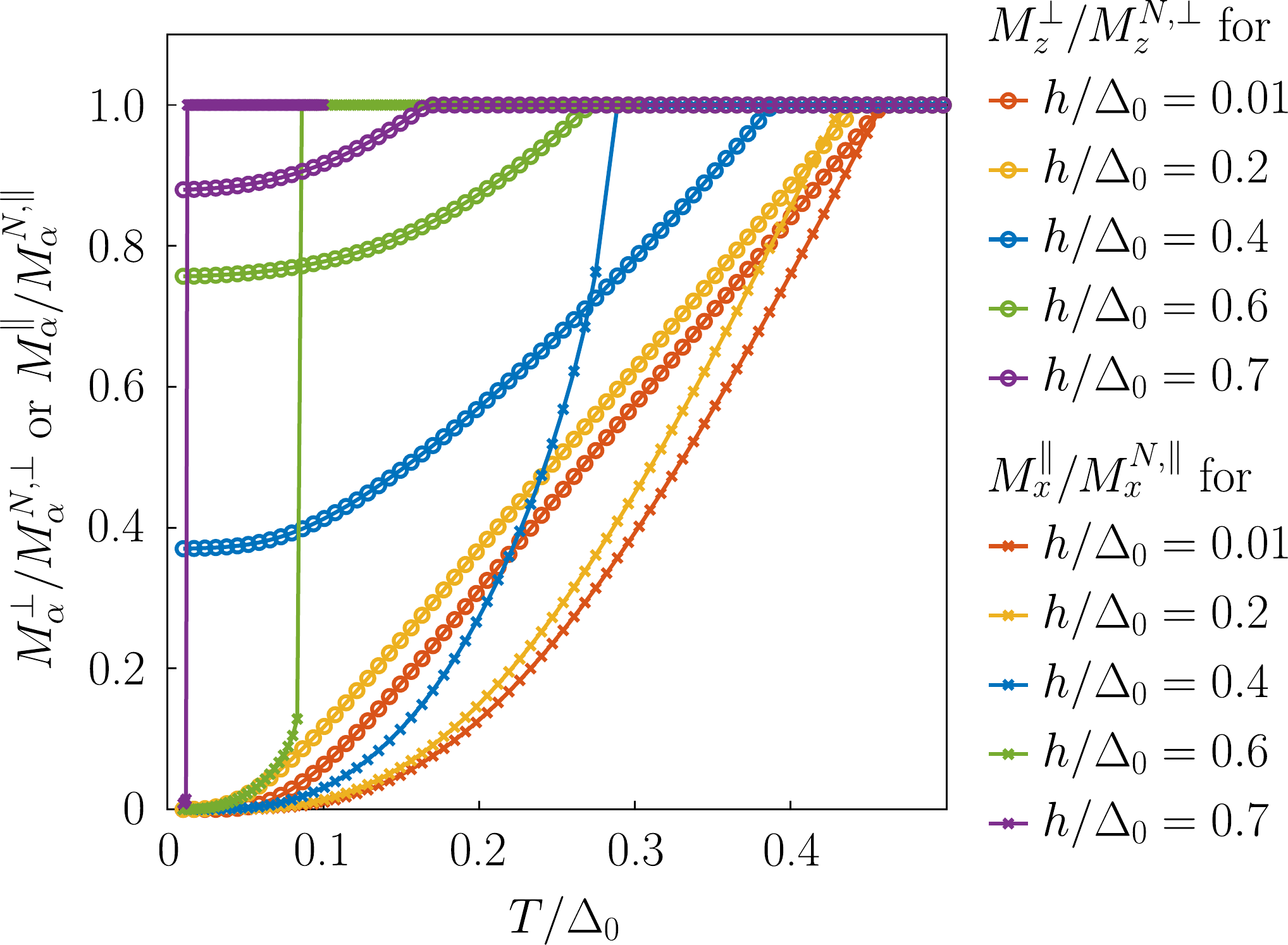}
    \caption{Superconductor's magnetization in an AM/S bilayer normalized by its normal-state value, $M^{\perp}_\alpha/M^{N,\perp}_\alpha$ ($M^{\parallel}_\alpha/M^{N,\parallel}_\alpha$) for $\alpha \in \{x, y, z$\}, is plotted as a function of temperature $T$ when an external magnetic field, $\vec{h}=h \, \hat{z}$ ($\vec{h}=h \, \hat{x}$), is applied out-of-plane (in-plane). Here, the altermagnetic exchange strength is taken to be $H_{AM}= 0.7\Delta_0$, and only the non-zero components of magnetization are plotted.}
    \label{Magnetization_vs_T_AMS}
\end{figure}

\begin{figure}[tb]
    \includegraphics[width=\linewidth]{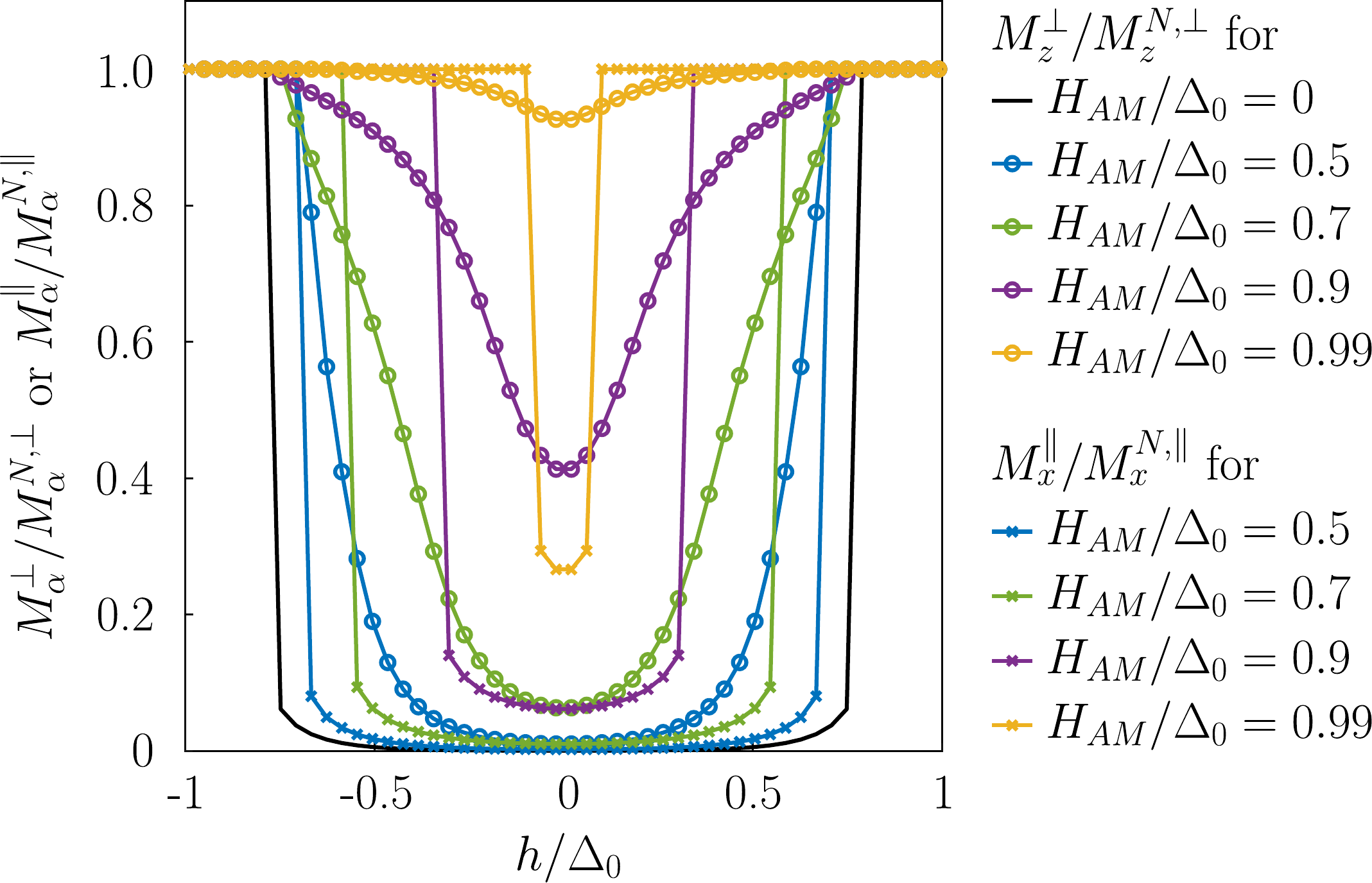}
    \caption{Superconductor's magnetization in an AM/S bilayer normalized by its a normal-state value, $M^{\perp}_\alpha/M^{N,\perp}_\alpha$ ($M^{\parallel}_\alpha/M^{N,\parallel}_\alpha$) for $\alpha \in \{x, y, z$\}, is plotted as a function of $h$, the magnitude of applied field $\vec{h}=h\, \hat{z}$ ($\vec{h}=h\, \hat{x}$) in the out-of-plane (in-plane) direction. Here, only the non-zero components of magnetization are plotted, and the temperature is $T=0.1\Delta_0$.}
    \label{Magnetization_vs_h_AMS}
\end{figure}

Fig.~\ref{Magnetization_vs_T_AMS} shows the temperature dependence of the superconductor's magnetization in an AM/S bilayer for different strengths of applied magnetic fields in the in-plane and out-of-plane directions. We observe that the magnetization $M_z^{\perp}/M_z^{N,\perp}$, when the applied field is out-of-plane, behaves similar to the linear susceptibility $\chi_{\perp}/\chi_{\perp}^N$ (red curve with circle markers in Fig.~\ref{Magnetization_vs_T_AMS}) for a weak applied field, whereas it becomes non-zero at zero-temperature for stronger fields. In presence of an out-of-plane applied field, the effective spin-splitting field is $\vec{H}_{\text{eff}}(\phi)=[H_{AM} \cos(2\phi)+h]\hat{z}$. When $H_{AM}+h>\Delta_0$, the superconducting gap in the DOS [Fig.~\ref{plot_DOS_and_PhaseDiagram}(a)] closes at zero-temperature making quasiparticles available for magnetization. However, one can see that it is possible to have $|\vec{H}_{\text{eff}}(\phi)|=\sqrt{H_{AM}^2\cos^2(2\phi)+h^2}<\Delta_0$ (for all $\phi$) for the same magnitude of the external field if it is in-plane, which provides a qualitative difference in the magnetization behavior at low temperatures, see the green and violet lines of Fig.~\ref{Magnetization_vs_T_AMS}.

The magnetization $M_x^{\parallel}/M_x^{N,\parallel}$ curves for an in-plane magnetic field in Fig.~\ref{Magnetization_vs_T_AMS}
behave similar to that of an isolated superconductor with a modified superconducting critical temperature $T_c$. This is because the AM's intrinsic exchange field is out-of-plane and does not affect the in-plane magnetization.

The suppression of $T_c$ of an AM/S is stronger when the magnetic field is applied in-plane (Fig.~\ref{Magnetization_vs_T_AMS}). This is because the magnetic field gets partially (in a part of the $\vec{k}$ space) screened by the AM's intrinsic exchange field when applied out-of-plane, which does not happen in the in-plane case.

Fig.~\ref{Magnetization_vs_h_AMS} shows the magnetization of the superconductor in an AM/S bilayer as a function of in-plane and out-of-plane applied magnetic field at $T=0.1\Delta_0$. In AM/S, the superconductor's magnetization, $M^{\parallel}_x/M^{N,\parallel}_x$, in an applied in-plane magnetic field behaves similar to an isolated superconductor (black curve in Fig.~\ref{Magnetization_vs_h_AMS}). This is consistent with Fig.~\ref{Magnetization_vs_T_AMS} and Fig.~\ref{LinearSusceptibility_vs_T_AMS} which show that the AM does not have a direct effect on the superconductor's in-plane magnetization. However, the out-of-plane magnetization $M^{\perp}_z/M^{N,\perp}_z$ curve shows a very different trend as the out-of-plane applied field interferes with the d-wave nature of the intrinsic exchange field of AM/S. Also, the magnetization in the out-of-plane case is always larger than that in the in-plane case. This is because the maximum magnitude of the exchange field in the out-of-plane case, $H_{AM}+h$, is larger than that in the in-plane case, $\sqrt{H_{AM}^2+h^2}$.

In summary, we applied the quasiclassical Green's function approach to study a few properties of the AM/S bilayer. To this end, we calculated the DOS for such a system and demonstrated that it resembles the DOS of an FM/S bilayer averaged over the varying exchange field. We plotted the phase diagram and the specific heat, which differ from the ferromagnetic case by the absence of first-order transitions for higher exchange fields. Another feature that can differentiate and help characterize such a bilayer is its response to an external magnetic field. The spin susceptibility shows strong anisotropy, specifically in-plane susceptibility being qualitatively similar to the case of an isolated superconductor, while the out-of-plane field reveals the d-wave nature of the magnetism in AM. These results are essential to understand the properties of new hybrid materials constituting AMs and superconductors. Moreover, the unique features of the AM/S bilayer should allow for experimental verification if the hybrid structure does possess both altermagnetic and superconducting properties or if the proximity effect suppresses either of them. 

\section*{Acknowledgments}
S.C. and A.K. acknowledge financial support from the Spanish Ministry for Science and Innovation -- AEI Grant CEX2018-000805-M (through the ``Maria de Maeztu'' Programme for Units of Excellence in R\&D) and grant RYC2021-031063-I funded by MCIN/AEI/10.13039/501100011033 and ``European Union Next Generation EU/PRTR''. S.C. also acknowledges Spanish MICINN (Grants Nos. PID2019-109539GB-C43, TED2021-131323B-I00 \& PID2022-141712NB-C21), Generalitat Valenciana through Programa Prometeo (2021/017), and the computational resources provided by Centro de Computación Científica of the Universidad Autónoma de Madrid. S.C. and W.B. acknowledge support by the Deutsche Forschungsgemeinschaft (DFG; German Research Foundation) through SFB 1432 (project ID 425217212) and and SPP2244 (project ID 417034116).

\section*{Data Availability Statement}
The data that support the findings of this study are available within the article and its supplementary material.

\appendix*

\section{\label{PHASE DIAGRAM OF FM/S} Phase diagram of FM/S}
\begin{figure}[tb]
    \includegraphics[width=\linewidth]{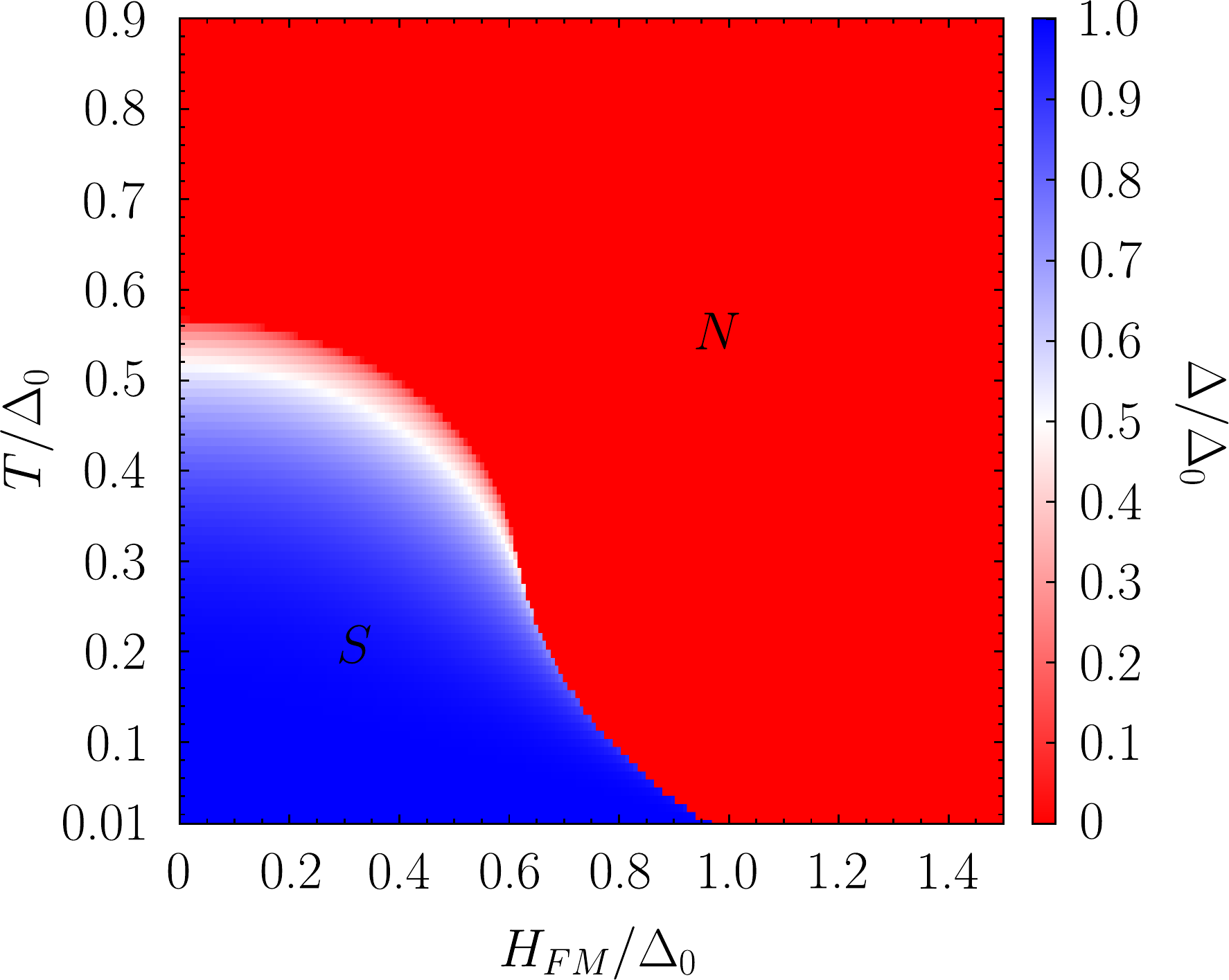}
    \caption{Phase diagram of FM/S obtained by plotting its superconducting order parameter $\Delta$ for different values of temperature $T$ and ferromagnetic exchange field $H_{FM}$.}
    \label{DeltaPhaseDiagram_FMS}
\end{figure}
In this section, we present the superconducting phase diagram of a thin film FM/S bilayer. We solve Eq.~\ref{Eilenberger1} for $\vec{H}_{\text{eff}}(\hat{k})=H_{FM}\hat{z}$, where $H_{FM}$ is the magnitude of uniform exchange field imparted by the insulating FM on the electrons of the superconductor. The phase diagram obtained by calculating $\Delta$ self-consistently is plotted in Fig.~\ref{DeltaPhaseDiagram_FMS}. We observe that the S-N transition with increasing temperature is second-order for smaller values of $H_{FM}$ and first-order for larger $H_{FM}$, which qualitatively agrees with a rigorous analysis done by comparing the free energies involved~\cite{Maki1964, Sarma1963}. In Fig.~\ref{DeltaPhaseDiagram_FMS}, the color changes from blue to white to red with increasing temperature for lower values of $H_{FM}$. The absence of the white region at higher $H_{FM}$ values indicates first-order S-N transition. However, in the case of AM/S, S-N transition with increasing temperature always has the white region indicating second-order transition for all values of $H_{AM}$ [Fig.~\ref{plot_DOS_and_PhaseDiagram}(b)].
\bibliography{AMS.bib}

\end{document}